# Russo-Ukrainian war disinformation detection in suspicious Telegram channels

Anton Bazdyrev*1*

*1 National Technical University of Ukraine "Igor Sikorsky Kyiv Polytechnic Institute", Kyiv, 03056, Ukraine*

**Abstract**
The paper proposes an advanced approach for identifying disinformation on Telegram channels related to the Russo-Ukrainian conflict, utilizing state-of-the-art (SOTA) deep learning techniques and transfer learning. Traditional methods of disinformation detection, often relying on manual verification or rule-based systems, are increasingly inadequate in the face of rapidly evolving propaganda tactics and the massive volume of data generated daily. To address these challenges, the proposed system employs deep learning algorithms, including LLM models, which are fine-tuned on a custom dataset encompassing verified disinformation and legitimate content. The paper's findings indicate that this approach significantly outperforms traditional machine learning techniques, offering enhanced contextual understanding and adaptability to emerging disinformation strategies.

## 1. Introduction

The Russo-Ukrainian war has been fought on the ground and in the digital realm, where disinformation and propaganda have become critical tools of influence. Telegram's anonymity and widespread use in military conflict zones have turned it into a go-to platform for spreading disinformation, especially via channels that offer real-time updates from the frontlines. The spread of manipulated narratives and misleading information poses serious threats, including undermining public trust, destabilizing societies, and manipulating international perceptions.

Traditional methods of disinformation detection usually rely on manual verification and rule-based systems and have proven inadequate in addressing the evolving tactics employed by propagandists. The dynamic nature of content, coupled with the sheer volume of data generated daily, calls for more scalable solutions. Recent advancements in Deep Learning offer promising avenues to enhance the detection capabilities in this domain.

This paper introduces an automated detection system for identifying disinformation within suspicious Telegram channels about the Russo-Ukrainian conflict. Leveraging the power of DL algorithms and transfer learning techniques, the system is designed to analyze textual content, identify patterns indicative of misinformation, and adapt to emerging disinformation strategies. Furthermore, the integration of more deep models enables a nuanced understanding of context, semantics, and linguistic subtleties that are often exploited in deceptive narratives.

A critical component of our approach involves processing a custom dataset encompassing verified disinformation instances and legitimate content from Telegram channels. This dataset serves as the training ground for our models, ensuring they capture the multifaceted nature of disinformation prevalent in the Russo-Ukrainian war context. Preliminary results demonstrate the efficacy of our system in identifying disinformation, boasting precision, and recall rates. These findings underscore the potential of integrating SOTA AI technologies to fortify the digital information landscape against malicious incursions.

## 2. Related works

### 2.1. Background

Detecting propaganda and disinformation in media is an evolving field, especially critical during periods of conflict. Propaganda is strategically designed to manipulate public opinion by promoting specific agendas, often through emotionally charged or misleading content. The paper "Fine-Grained Analysis of Propaganda in News Articles" [1] addresses the challenge of detecting propaganda within textual content more granularly than previously attempted. Traditional approaches typically labeled entire articles from a propagandistic outlet as propaganda, leading to noisy gold labels compromising the quality of learning systems trained on such data. This research introduced a novel approach by developing a corpus annotated at the fragment level with eighteen distinct propaganda techniques, enabling the identification of specific propagandistic elements within texts.

The relevance of this work [1] to disinformation detection is significant but indirect. While propaganda is a vital tool within the broader disinformation framework, it does not encompass the entirety of disinformation activities. Disinformation can be subtle, sometimes devoid of overt manipulative techniques, and may manifest in forms that do not fit neatly into predefined propaganda categories. This limitation underscores the necessity of expanding beyond fine-grained propaganda analysis to incorporate a broader range of disinformation strategies, particularly those that leverage platforms like Telegram, which are prominent in the Russo-Ukrainian conflict.

### 2.2. Relevant data and disinformation modeling

The rise of disinformation surrounding the Russo-Ukrainian war is increasingly fueled by advancements in generative technologies like Large Language Models (LLMs). A recent study [2] examines how these technologies impact disinformation detection. LLMs, such as ChatGPT, are incredibly skilled at producing human-like text, which disinformation spreaders can use to create convincing and deceptive content. The study highlights three critical research questions: the effectiveness of current detection techniques against LLM-generated disinformation, the potential for LLMs to be used as a defense mechanism, and the exploration of novel strategies to combat these emerging threats. This paper is particularly pertinent to the detection of disinformation in Telegram channels during the Russo-Ukrainian war, where LLMs could be used to generate content that mimics authentic posts, making it difficult to distinguish between legitimate and deceptive information. The research underscores the need for continuous adaptation of detection systems to counter the evolving tactics of disinformation agents. It suggests that leveraging the capabilities of LLMs themselves might provide a dual-use solution—both as a threat and as a tool for detection.

Effective disinformation detection heavily relies on the quality and diversity of datasets used for model training. The research [3] provides an in-depth review of the existing datasets for fake news detection, highlighting the importance of the data quality in the effectiveness of detection models. The survey emphasizes the importance of diverse labeling systems and addresses the ethical issues associated with dataset creation and use. The paper also discusses the potential biases in datasets, which can significantly impact the performance of models, particularly in conflict zones where information is highly polarized.

There are also qualitative studies, e.g., in paper [4], the authors address the specific challenge of detecting pro-Russian propaganda within Telegram channels in Odesa. The study is highly relevant to our research because it provides a concrete example of how qualitative analysis, for example frame analysis, that can be applied to understand the narratives and frames used in disinformation campaigns. The study identified keyframes like "inconvenience" and "persecution," which are used to promote pro-Russian narratives while downplaying Russian aggression subtly.

The findings from this paper suggest that frame analysis can be an effective tool for identifying the underlying themes and narratives in Telegram posts, which can then be used to enhance automated detection systems. By focusing on specific frames, such as those promoting regional separation or language persecution, it is possible to develop targeted algorithms that detect these narratives with higher accuracy. Moreover, the study's emphasis on the regional context of Odesa highlights the importance of considering local dynamics when developing detection systems, as disinformation strategies can vary significantly across different regions.

### 2.3. Research gap

Though progress has been made in detecting disinformation and propaganda on various media platforms. There are still important gaps that need immediate attention, especially the spread of disinformation on Telegram channels. While current research provides valuable insights into identifying propaganda and disinformation, the lack of fully automated systems to recognize and flag harmful content in real-time is a pressing issue. Current methods often require significant manual intervention or are limited to post-hoc analysis, which is inadequate in the fast-paced environment of Telegram, where disinformation can spread rapidly.

Secondly, even current SOTA detection systems cannot evaluate the potential risk level of disinformation. Disinformation can have varying levels of impact, with particular messages posing a more immediate threat to public safety, social stability, or national security. An adequate system should not only identify disinformation but also provide a measurable assessment of its level of danger, allowing for timely interventions for the most harmful content.

These gaps highlight the necessity of developing an automated detection system that not only identifies disinformation in real-time but also assesses its danger level and functions at the granular message level, tailored specifically for the dynamic environment of Telegram channels in the Russo-Ukrainian war.

## 3. Dataset and Evaluation

### 3.1. Main dataset

The original dataset was provided by the Disinformation Detection Challenge Kaggle competition by AI HOUSE x Mantis Analytics [5, 6]. The full dataset consists of approximately 1500 posts from 56 different Telegram channels, where each post is labeled with a Suspicious_Level target by human annotators. The Kaggle train part consists of posts from July 2023, and the public/private test part consists of posts from September 2023. Some examples from the dataset are provided in Figure 1, and the distribution of classes is in Figure 2.

Possible suspicious levels:

1. likely not suspicious
2. doubtful, potentially needed to be checked
3. surely suspicious disinformation

In the manual pipeline, the annotator has a list of tips that helps them make decisions:
- From war correspondents and the general situation at the front – 1
- Photos/videos of the attack - 2 (sometimes there are fake videos, but most of them cannot be verified)
- Photos/videos of the attack, where posts mention leopards or other Western weapons - 3; there were fakes about leopards already, Russia is actively working to discredit Western equipment
- Calls to support money raising – 1
- US statements about the new aid package – 1
- Video of mobilization in Ukraine – 3
- Interviews from Volyna about Azovstal and American soldiers – 3

- Lavrov will meet with someone – 1
- LGBTQ+ - 2/3

```
Not suspicious:

По атакам в районе Керчи и в Ростовской области.

1. Движение по Крымскому мосту в обе стороны после временного закрытия моста возобновлено.
ПВО-шники четко отработали цель. Сейчас устанавливается ее тип.

2. В Ростовской области ПВО отрабатывала в направлении ЛНР. Поражены 2 цели (тип неизвестен пока что), о каких
-либо прилетах, жертвах и разрушениях не сообщалось.
-------------------------------
Suspicious:

!!🇺🇦📛Пехоту для "мясных штурмов" силой хватают на улицах Украины и волокут на убой на фронт
Новые кадры из Одессы, в регионе продолжают появляться видео с грубым насильственным вручением повесток. Из-за
этого в области даже на днях сменили военкома, но это не помогло, судя по всему.
Массовые рейды продолжаются по всей Украине, ВСУ пытаются восполнить свои огромные потери на фронте.
```

**Figure 1**: Example of not suspicious (valid info) and suspicious (disinformation)

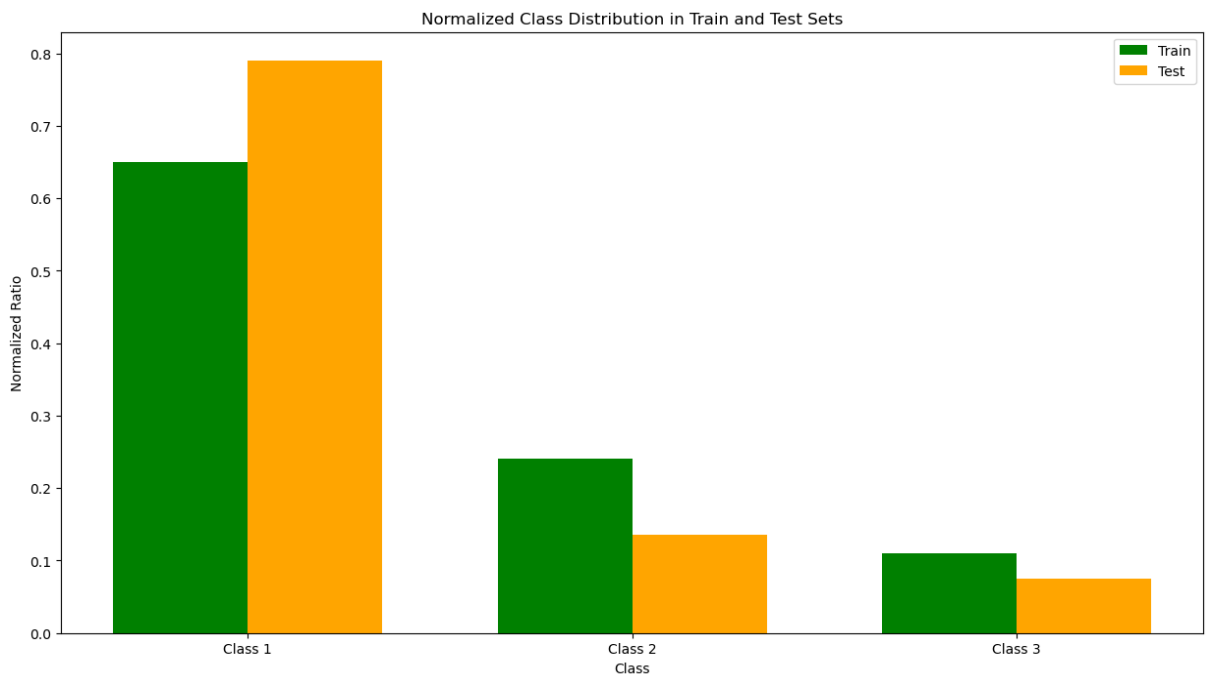

**Figure 2**: Distribution of classes in train and test

### 3.2. Additional dataset

For our research, we additionally collected all posts from the telegram channels of the main dataset between July and October 2023, a total of 250k records.

### 3.3. Evaluation approach

The primary metric for the evaluation is F1 with macro average [7]. It is particularly beneficial for multiclass tasks with class imbalance because it provides a balanced evaluation of model performance across all classes, regardless of their individual sizes. Even if a model performs well on the majority class, poor performance on a minority class will significantly lower the F1 macro score, highlighting the model's weaknesses in those classes. We also use 5-fold cross-validation with shuffle for local validation purposes.

## 4. Experiments

In this section, we will review all of our various experiments conducted for this study. Table 1 and Table 2 present all aggregated final metric values.

### 4.1. Baseline ML approach

Our baseline ML solution is TF-IDF method for vectorization stacked with logistic regression [8]. This technique relies on statistical text representation combined with linear classification to deliver satisfactory results. The key idea is to convert text data into numerical features that a statistical model can process. TF-IDF vectorization accomplishes this by creating a matrix of numerical features, reflecting both the frequency of words in individual documents and their significance within the broader collection of text. After transforming the text into vectorized representation features, logistic regression is then used to classify the data. We used 10k max features in TF-IDF vectorizer and C=1.0 for logistic regression l2 regularization. This method did not work well for this task. It seems that it requires a much deeper contextual understanding of the input text.

### 4.2. Classical DL approach

Our classical DL solution is xlm-roberta-large (355M parameters) [9, 10] pretrained multilingual model with linear head finetuning on top of it. This method generally performed better than traditional ML because of its superior contextual representation of the text. However, in some experiments, it appeared that the model still struggles to understand the context of events in the ongoing Russo-Ukrainian war, leading to weird results. Additionally, it is almost impossible to achieve good results when fine-tuning the entire model with the backbone, rather than just the classifier head, because of almost immediate overfitting.

### 4.3. Transfer learning approach with additional pre-training

The idea of this approach is very similar to the previous one, with the key difference being that we performed the entire pre-training of xlm-roberta-large (355M) on the MLM task like in the original paper [10] ourselves using additional posts from Telegram, as mentioned earlier in the "Additional dataset" chapter. The results were the best among all the experiments that we tried. This can be qualitatively explained by the fact that the domain of warfare is very specific. Existing pre-trained models often have little knowledge in this domain area, so to fully leverage transfer learning, it is necessary to perform pre-training from scratch on the relevant data. The pre-training step here is very resource-intensive. For our additional dataset of 250k samples and 5 epochs, it took around 8 hours on the 4xV100 GPU cluster. For fine-tuning, we still tune only the final classification layer to avoid overfitting. This ensures the model has an excellent contextual understanding of the input text, and has strong generalization ability.

### 4.4. LLM prompt-engineering

Although proprietary models and models larger than 1B parameters were prohibited in the original Kaggle competition, it makes sense from a research perspective to explore different techniques for working with LLMs. The simplest option is to use SOTA proprietary models with prompt engineering. In this experiment, we used GPT-4-1106 as a starting point. After experimenting with different prompts, we found the optimal prompt to be: "You are an expert in news analysis and disinformation. Your task is to classify each message into one of three possible categories (Suspicious_Level): <list of categories>. During the manual process of annotation, the annotation tool contains a list of tips that help you make decisions: <list of tips>". Where <list of categories> and <list of tips > were used with respect to the Main dataset chapter. The final results were quite contradictory. This can be qualitatively explained by the fact that the original dataset is very imbalanced. With prompt engineering, it is difficult to alter the class distribution in the model's output as generated text, which negatively impacts the macro F1 score.

### 4.5. LLM LoRA fine-tuning

In this approach, we aim to fully leverage both the power of state-of-the-art LLMs and fine-tuning to achieve the best possible results. Therefore, we use the following configuration:

Therefore, we use the following configuration:
- Gemma2-9b-it [11, 12] LLM model
- LoRA [13] tuning with o_proj, v_proj, q_proj, k_proj target modules; rank=32; alpha=16
- 12 first layers freeze to reduce overfitting
- batch_size=4; learning_rate=2e-4; scheduler=cosine

This approach worked quite well and was relatively fast (10 minutes of training on 1xV100) because the current pre-trained model is already quite intelligent, and LoRA fine-tuning only trains 0.3% of all the parameters of the original model. However, the results were worse than the option with full pre-training using XLM-RoBERTa, likely because the pre-training data for Gemma2 either did not include Telegram data on the war or was partially censored, leading to worse out-of-the-box performance. Conducting a full pre-training of Gemma2 on Telegram data is not possible for us due to the enormous computational resources required.

### 4.6. Metrics and final multi-class results

Table 1 presents the results of all experiments, with an added row labeled "Naïve Random" which refers to predicting the most frequent class. This allows us to see how much better our ML solutions perform compared to the baseline. The "5-Fold CV" column shows the metric values on the validation splits of the public training dataset, while the "Kaggle Public" and "Kaggle Private" columns display the metric values on the public and private subsets of the Kaggle benchmark, respectively. These columns enable comparison of our solutions with those of other participants in the competition.

It is quite noticeable that the metric results in the three columns differ significantly from each other. This can be explained by the obvious temporal factors in the data, as different data subsets contain posts from different dates, which affects the topics and vocabulary of the posts and introduces a domain shift in the data.

Although our results are only ranked 2nd out of 38 for this benchmark on the private leaderboard, they are stable and consistent across different test data splits, unlike the top private leaderboard entry, which does not appear in the top 5 on the public leaderboard and vice versa. This consistency is beneficial for potential deployment in production, as it suggests that the model's performance on new, previously unseen data is likely to be more stable.

**Table 1**
**Comparative results of multiclass experiments**

| Experiment | 5-Fold CV | Kaggle Public | Kaggle Private |
| --- | --- | --- | --- |
| Naïve Random | 0.26 | 0.29 | 0.29 |
| TF-IDF + LogisticReg | 0.43 | 0.32 | 0.32 |
| XLM-RoBERTa-LARGE | 0.55 | 0.44 | 0.42 |
| XLM-RoBERTa-LARGE full pretrained (250k) | ***0.62*** | ***0.5*** | ***0.47*** |
| GPT-4-1106 prompt engineering | 0.52 | 0.43 | 0.38 |
| Gemma2-9b-it LoRA fine-tuning | 0.58 | 0.45 | 0.43 |

We also present the top Public and Private Kaggle leaderboards in Figure 3 and Figure 4, respectively.

**Figure 3**: Top 5 Kaggle Public Leaderboard

**Figure 4**: Top 5 Kaggle Private Leaderboard

## 4.7. The binarized simplification of the problem

Both from a human labeling perspective and from a model performance perspective, it can be challenging to distinguish between different suspicious levels in texts among those that are potentially suspicious. According to the Figure 2, both suspicious classes 2 and 3 could be considered as minority classes, where class 3 consists of rather more suspicious posts than class 2. We can consider a binary version of this problem by simply dividing all posts into suspicious and non-suspicious, where non-suspicious includes class 1 and suspicious includes 2 and 3. For a binary version of the problem we use ROC-AUC, Precision-Recall AUC (AUC-PR), binary F1 score and 5-folds CV similar to the previous section.

**Table 2**
**Comparative results of binary experiments**

| Experiment | F1 | ROC-AUC | AUC-PR |
| --- | --- | --- | --- |
| TF-IDF + LogisticReg | 0.765 | 0.928 | 0.897 |
| XLM-RoBERTa-LARGE | 0.852 | 0.946 | 0.911 |
| XLM-RoBERTa-LARGE full pretrained (250k) | *0.891* | *0.960* | *0.919* |
| Gemma2-9b-it LoRA fine-tuning | 0.880 | 0.957 | 0.916 |

### 4.8. Inference and practical use aspects

In terms of accuracy by metrics in both binary and multiclass cases, models based on fine-tuned transformers work significantly better than classical methods and than third-party APIs of pre-trained LLMs. From the hardware optimization side, the current version of the best solution for metrics (XLM-RoBERTa-LARGE with full pre-training) could work in CPU environments - approximately 1 request per second on 4xCPU 16 GB RAM machine with 8-bit quantization. If we want to use this on a large scale while being gpu-poor, then the most optimal solution would be to use a binary version TF-IDF + LogisticReg model with a moderate threshold as an input filter and then perform more accurate classification using the XLM-RoBERTa model in order to reduce number of posts classified by the compute-intensive XLM-RoBERTa model, but this approach could lead to evaluation metrics degradation.

## 5. Conclusion

We have obtained very promising results from our experiments with employing an automatic detection system for identifying disinformation in Telegram channels discussing the Russo-Ukrainian conflict. As we performed pre-training on a custom dataset as part of our transfer learning, we found that other approaches such as traditional machine learning techniques and deep learning models were outperformed on different evaluation metrics. Despite these findings, there are still challenges to overcome before this system can be used in real-world scenarios. This stage of pre-training is highly dependent on GPU and requires substantial computational power; thus, scalability in this approach is limited. Moreover, tests conducted on different data subsets showed significant variations that could arise from changes over time in the dataset. Hence, this may undermine the system's utility in practical situations. These discrepancies must be addressed to adapt to changing conditions and enhance usability across diverse contexts.

## Acknowledgements

First and foremost, I want to thank my students from the Kaggle team Adam's sons: Andrii Shevtsov, Andrii Zhuravlov and Yevhen Herasimov, for their part in the Kaggle experiments. I am also grateful to Volodymyr Sydorskyi, Oleg Yaroshevskiy, Maxim Tereschenko and the Mantis Analytics team for preparing the dataset and organizing the Kaggle competition.